\documentclass[sigconf,screen,table,nonacm]{acmart}

\acmBooktitle{Proceedings of the 32nd ACM Symposium on the Foundations of Software Engineering (FSE '24), November 15--19, 2024, Porto de Galinhas, Brazil}

\usepackage{amsmath,amsfonts}
\usepackage{ifthen}
\usepackage{graphicx}
\usepackage{adjustbox}
\usepackage{textcomp}
\usepackage{xcolor}
\usepackage{siunitx}
\usepackage{pifont}
\usepackage{hyperref}
\usepackage[normalem]{ulem}
\usepackage{tcolorbox}
\usepackage{comment}
\usepackage{subcaption}
\usepackage{lscape}
\usepackage{listings}
\usepackage{hyperref}
\usepackage{xspace}
\usepackage{latexsym}
\usepackage{listings}
\usepackage[frozencache]{minted}
\setminted[java]{
xleftmargin=15pt,
framesep=2mm,
linenos=true,
breaklines=true,
tabsize=2,
encoding=utf8,
frame=none,
fontsize=\small,
escapeinside=|| 
}

\usepackage[textsize=tiny]{todonotes}
\presetkeys{todonotes}{fancyline}{}

\definecolor{excerptblue}{HTML}{33FFAC}
\definecolor{excerptgray}{HTML}{BEB8C6}

\newcommand{\dcircle}[1]{\ding{\numexpr171 + #1}}
\newcommand{\bcircle}[1]{\ding{\numexpr181 + #1}}

\newcommand{\al}{AndroLog\xspace}

\newcommand{\highlight}[1]{%
\begin{tcolorbox}[leftrule=1mm,rightrule=1mm,toprule=0mm,bottomrule=0mm,left=0pt,right=0pt,top=0pt,bottom=0pt, colback=gray!30, colframe=gray!90]
#1
\end{tcolorbox}%
}

\begin{document}

\title{\al: Android Instrumentation and Code Coverage Analysis}

\author{Jordan Samhi}
\orcid{0000-0001-6052-6184}
\affiliation{%
	\institution{CISPA Helmholtz Center for Information Security}
  	\city{Saarbr{\"u}cken}
	\country{Germany}
}
\email{jordan.samhi@cispa.de}

\author{Andreas Zeller}
\orcid{0000-0003-4719-8803}
\affiliation{%
	\institution{CISPA Helmholtz Center for Information Security}
  \city{Saarbr{\"u}cken}
	\country{Germany}
}
\email{zeller@cispa.de}

\begin{abstract}
Dynamic analysis has emerged as a pivotal technique for testing Android apps, enabling the detection of bugs, malicious code, and vulnerabilities. 
A key metric in evaluating the efficacy of tools employed by both research and practitioner communities for this purpose is \emph{code coverage.} 
Obtaining code coverage typically requires planting probes within apps to gather coverage data during runtime. 
Due to the general unavailability of source code to analysts, there is a necessity for instrumenting apps to insert these probes in black-box environments.
However, the tools available for such instrumentation are limited in their reliability and require intrusive changes interfering with apps' functionalities.

This paper introduces \emph{\al,} a novel tool developed on top of the \emph{Soot} framework, designed to provide \emph{fine-grained coverage information} at multiple levels, including class, methods, statements, and Android components. 
In contrast to existing tools, \al leaves the responsibility to test apps to analysts, and its motto is \emph{simplicity}. 
As demonstrated in this paper, \al can instrument up to 98\% of recent Android apps compared to existing tools with 79\% and 48\% respectively for COSMO and ACVTool.
\al also stands out for its potential for future enhancements to increase granularity on demand.
We make \al available to the community and provide a video demonstration of \al (see section~\ref{sec:data_availability}).
\end{abstract}

\maketitle

\section{Introduction}
\label{sec:introduction}

In recent years, analysts have identified numerous threats to Android security, even in official app markets like Google Play, making it easy for malware to reach millions of users.
As a result, ensuring the security and privacy of Android devices has become increasingly important.
Effective protection against these threats is paramount, as they can have severe consequences for individuals and organizations.
Researchers have developed various approaches to combat security and privacy threats in Android apps in response to these threats. 
These approaches include static analyses ~\cite{10.1109/NTMS.2016.7792435,doi:10.1155/2015/479174,10.1145/3510003.3512766,revisiting_shso}, dynamic analyses~\cite{van2013dynamic,10.1145/2592791.2592796}, and machine learning techniques~\cite{10.1109/EISIC.2012.34,10.1109/ICTAI.2013.53,10.1007/978-3-030-87839-9_4}.

Dynamic analysis is a prevalent technique in Android app analysis~\cite{google_protection}. 
Its widespread use is largely attributed to its effectiveness to uncover app security issues~\cite{10.1145/2491411.2491450,10.1145/2594368.2594390,7413692,Cui2023,10.1145/3238147.3240479,10.1145/3387903.3389310}.
A key challenge in dynamic analysis is determining the parts of the code being executed, a concept referred to as \emph{code coverage}~\cite{yang2013grey,tikir2002efficient}.
With access to source code, analysts can easily modify the code to add probes that track execution behaviors. 
But in reality, most apps are distributed in a "black-box" manner, i.e., the source code is inaccessible to analysts. 
This lack of access significantly complicates modifying the code to insert probes.
Hence, the ability to insert these probes in a black-box manner is paramount for analyzing apps on a large scale.

The most recent tools \emph{available} (most researchers developed ad-hoc techniques not publicly available) to instrument apps in a black-box manner are BBoxTester~\cite{7299958} (based on Emma~\cite{emma}), COSMO~\cite{9438584} (based on JaCoCo~\cite{jacoco}), and ACVTool~\cite{10.1145/3395042}.
These tools, however, come with a range of limitations and challenges.
Firstly, they are significantly outdated, with no maintenance for 8, 4, and 4 years, respectively for BBoxTester, ACVTool, and COSMO. 
This lack of updates renders them almost inapplicable for recent apps due to the rapid evolution of the Android platform.
Secondly, they require complex modifications to the apps (due to the use of existing frameworks to instrument Java programs, such as JaCoCo~\cite{jacoco} and Emma~\cite{emma}).
This involves adding new intrusive classes, Android components, resources, and structures in memory, and modifying the AndroidManifest.xml file with additional permissions, which can lead to unexpected or even altered behavior.
Thirdly, user experience with existing tools is cumbersome due to their complex setup and usage requirements (e.g., writing on the device's SD card).

\begin{figure}
    \centering
    \begin{adjustbox}{width=.8\columnwidth,center}
        \includegraphics[]{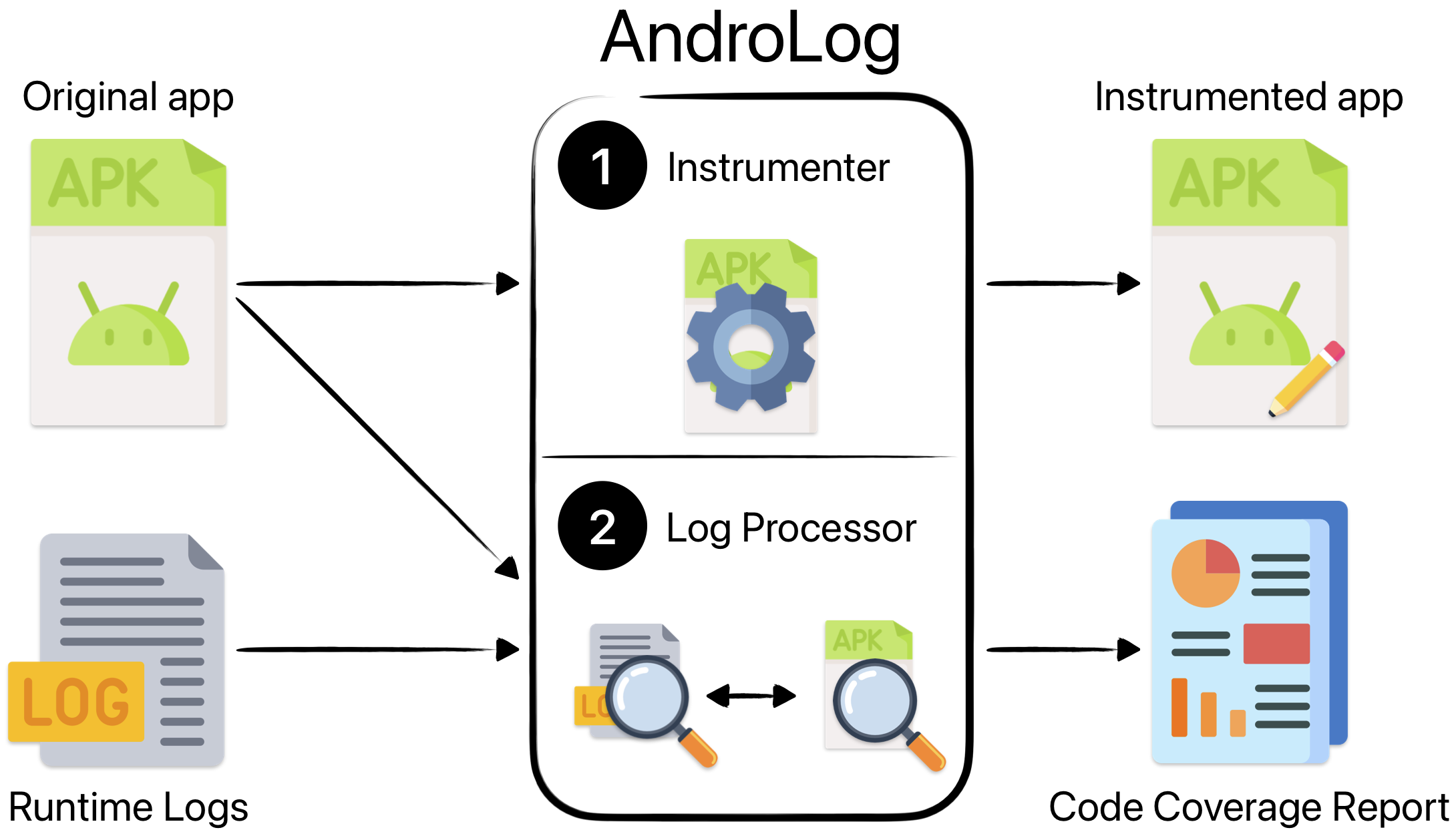}
    \end{adjustbox}
    \caption{\al's Architecture. The \al \emph{Instrumenter} \bcircle{1} takes an app and instruments it. the \al \emph{Log Processor} \bcircle{2} takes the app and the logs produced by the instrumented app to generate a code coverage report.}
    \label{fig:architecture}
\end{figure}

We introduce \al, a novel open-source framework for instrumenting Android apps in a black-box manner that overcomes all of the abovementioned deficiencies with a more efficient and simple solution.
Indeed, \al is conceptually designed with \emph{simplicity} and \emph{flexibility} in mind.
Most importantly, as instrumentation must not interfere with existing code, \al takes a streamlined approach by only adding log statements to monitor app behavior, with no new permissions, resources, Android components, or manifest modification.
This minimizes the risk of interfering with apps' original functionalities.
Additionally, \al:
\dcircle{1} is \emph{test-independent,} allowing analysts to choose whether to use \al for code coverage computation.
This approach simplifies the entire process and gives analysts more control over their testing and analysis procedures;
\dcircle{2} simplifies the instrumentation and code coverage computation process to just two main steps: (1)~providing an APK that is then instrumented into an APK$'$, and (2)~executing APK$'$ and feeding the resulting logs into \al to obtain a coverage report (note that the resulting logs is a valuable resource that can independently be used for any other downstream task);
\dcircle{3} relies on the Soot framework~\cite{10.5555/781995.782008}, ensuring its long-term viability and compatibility with newer apps (Soot is well tested and constantly maintained); and
\dcircle{4} allows for high customization for new levels of granularity as demanded by the evolving needs of dynamic analysis.

In summary, we make the following contributions:
\begin{itemize}
    \item We introduce \emph{\al,} a black-box instrumentation tool to compute code coverage of Android apps.
    \item \al works at various \emph{levels of granularity} (e.g., classes, methods, statements, and Android components) and can evolve on demand.
    \item We show \al's effectiveness in instrumenting the most recent apps compared to existing tools.
\end{itemize}

\highlight{
\al is conceived as a modern, efficient, easy-to-use, and adaptable tool, addressing the limitations of existing solutions and paving the way for more effective dynamic analysis and code coverage computation in Android apps.
}
\section{Architecture}
\label{sec:architecture}

\al's architecture is simple and straightforward, it is depicted in Figure~\ref{fig:architecture} and described hereafter.

\al is divided into two main components:
\bcircle{1} the first component is the instrumenter.
Its primary function is to process an APK, the standard file format for Android apps.
The instrumenter role involves adding log statements at various levels within the APK (as described in Section~\ref{sec:design}). 
The output of this phase is an instrumented APK', ready for testing; and
\bcircle{2} the second component is the log processor.
This module is designed to handle two inputs: the original APK file and the runtime logs collected by the analyst during the testing phase.
The responsibility of the log processor is to analyze these inputs and generate a code coverage report.

\begin{figure}
    \centering
    \begin{adjustbox}{width=.65\columnwidth,center}
        \includegraphics[]{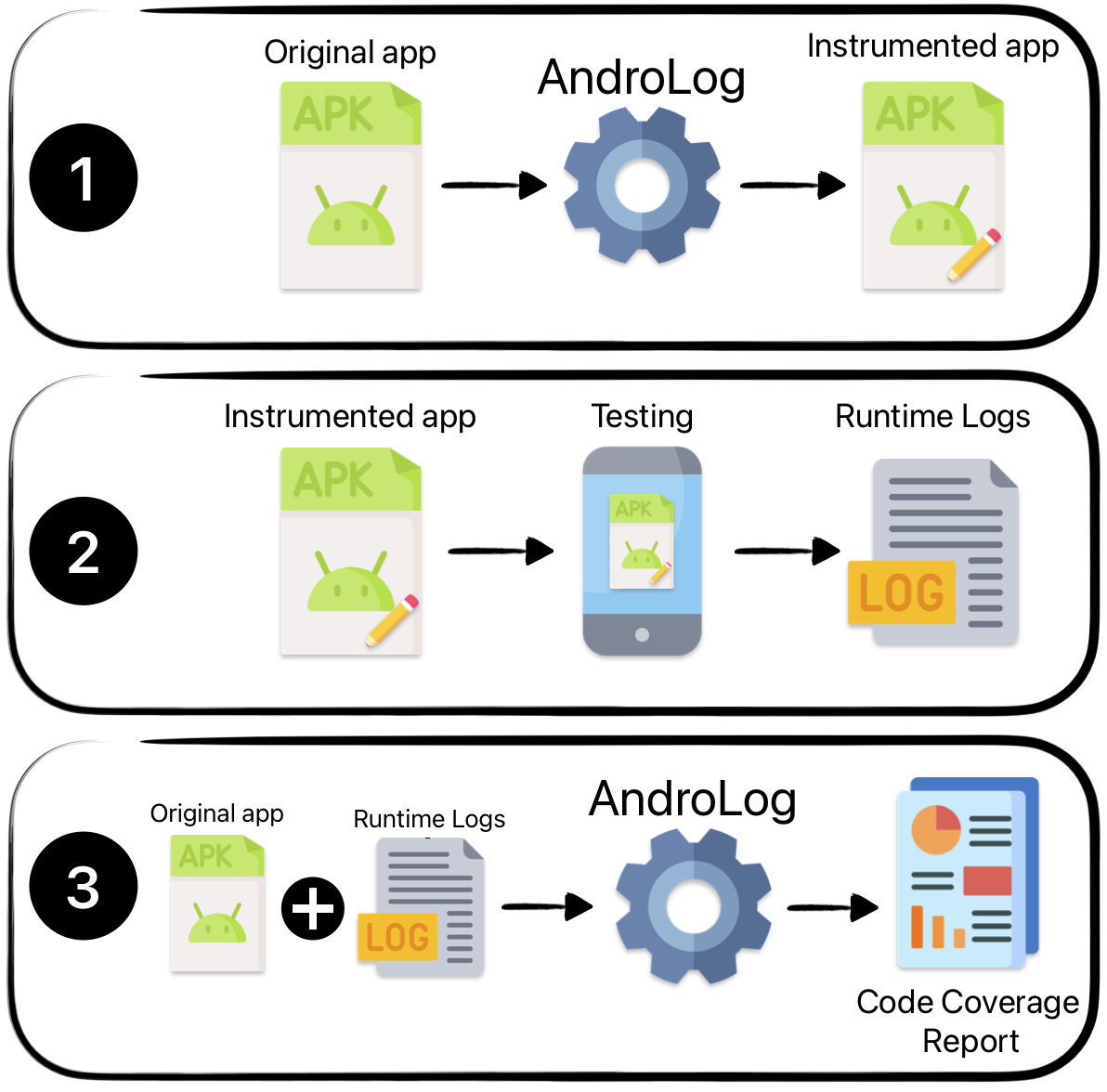}
    \end{adjustbox}
    \caption{\al's Workflow. \bcircle{1} The analyst uses \al to instrument an app. \bcircle{2} The analyst tests the instrumented app and collects runtime logs. \bcircle{3} The analyst uses \al to generate the code coverage report.}
    \label{fig:workflow}
\end{figure}

\noindent
\textbf{\al's Workflow.}
The operation of \al is encapsulated in a straightforward, three-step workflow, as depicted in Fig.~\ref{fig:workflow}:
\begin{enumerate}
    \item \textbf{Intrumentation phase.} The workflow begins with the analysts using \al to instrument the original app in order to add probes at various levels of granularity.
    \item \textbf{Testing phase.} The next step involves the execution/testing of this modified app on a device or an emulator to collect runtime logs. This is the analyst's responsibility.
    By design, \al has no role in this phase.
    \item \textbf{Code Coverage phase.} The final step is the generation of the code coverage report. Here, the analyst provides \al with two inputs: the original APK and the log report. \al then computes the code coverage report and displays it to the analyst.
\end{enumerate}

\noindent
\textbf{Functionalities}
\al allows the analyst to
\dcircle{1} sets the output folder of the instrumented app;
\dcircle{2} set the log identifier to parse the logs easily;
\dcircle{3} choose what to measure in terms of code coverage (i.e., classes, methods, statements, Android components, etc.); and
\dcircle{4} choose whether to include libraries for computing code coverage.

The interested reader can check our Github page for more information and glance at our online demonstration (see section~\ref{sec:data_availability}).

\section{Design}
\label{sec:design}

\begin{figure*}
    \centering
    \begin{adjustbox}{width=1.85\columnwidth,center}
        \includegraphics[]{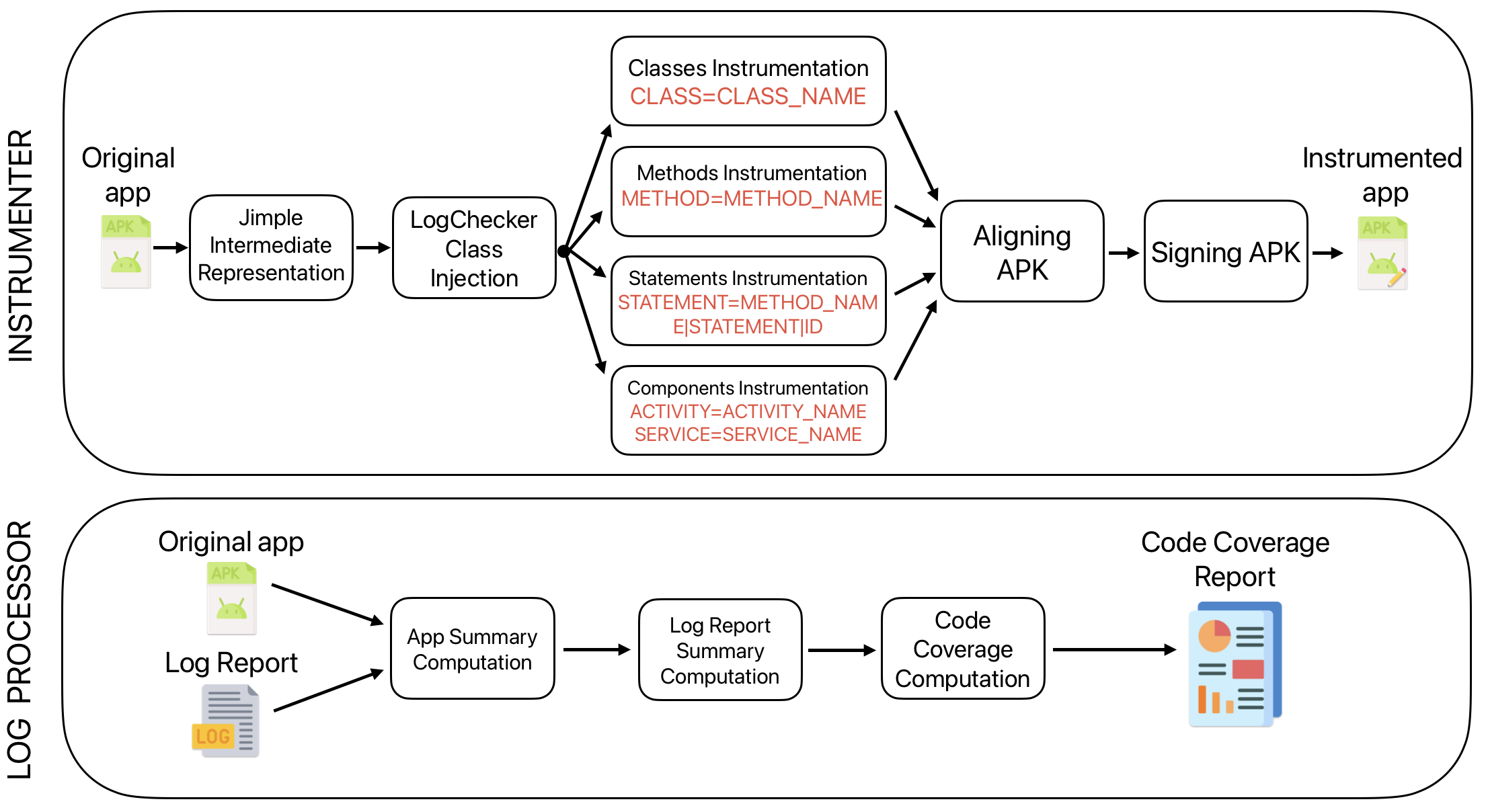}
    \end{adjustbox}
    \caption{\al's Design. \emph{Instrumenter:} \al transforms the Dalvik bytecode of an app into Jimple and instruments it. Then, \al aligns and signs the app which is ready for testing. \emph{Log Processor:} \al takes the original app (i.e., the non-instrumented version) and the execution logs to produce the code coverage report.}
    \label{fig:design}
\end{figure*}

In this section, we give the implementation details of \al.
As described in the previous section, \al is made of two main components:
\dcircle{1} the \emph{Instrumenter}; and
\dcircle{2} the \emph{Log Processor}.
The overview of \al's design is depicted in Figure~\ref{fig:design}; we describe both components hereafter.

\begin{listing}[h]
\begin{minted}{java}
public class LogCheckerClass {
    private static final Set LOGS = new HashSet();

    public static void log(String log, String tag) {
        if (LOGS.contains(log)) {
            return;
        }
        Log.d(tag, log);
        LOGS.add(log);
    }
}
\end{minted}
    \caption{LogCheckerClass injected into apps.}
    \label{code:logcheckerclass}
\end{listing}

\subsection{Instrumenter Component}

\al first transforms the Dalvik bytecode into the Jimple IR~\cite{vallee1998jimple} using the Soot framework.
Then, before the logs are inserted, and to ensure that the execution logs of an app are not overloaded and the app is not slowed down, \al injects a special class \texttt{LogCheckerClass} that keeps track of elements already logged and only prints a log if it was not previously logged. 
Listing~\ref{code:logcheckerclass} depicts the class injected.
Any instrumentation, described hereafter, then simply involves adding a call to the method \texttt{LogCheckerClass.log()} (Line~4 in Listing~\ref{code:logcheckerclass}) as depicted in Listing~\ref{code:instrumentation}.

\begin{listing}[h]
\begin{minted}{java}
public void run() {
+  LogCheckerClass.log("METHOD=<e: void run()>", "TAG");
    [...]
}
\end{minted}
    \caption{Example of the instrumentation of a method.}
    \label{code:instrumentation}
\end{listing}

The instrumentation process is then divided into four distinct phases, each targeting different app components (activated by the analyst):
\dcircle{1} In the first phase, \al iterates over all the classes within the app. 
For each class, it instruments both their constructors (identified as \texttt{void <init>()} in Jimple) and static constructors (\texttt{void <clinit>()} in Jimple). 
The rationale behind this is that any class usage, whether through instantiation or static access, involves these constructors. 
The instrumentation process involves adding a log statement as the first executed statement in these constructors. 
For example, a log statement would be: \texttt{LogCheckerClass.log\-("CLASS=com.example.MyClass", "ANDROLOG")}.
This ensures that any interaction with the class is logged;
\dcircle{2} The second phase involves iterating over all methods in the app. 
For each method, \al inserts a log statement at the very beginning of the code (see Listing~\ref{code:instrumentation}).
A method log statement would be: \texttt{LogCheckerClass.\-log("METHOD=com.example.Class.\-foo()", "ANDROLOG")}.
This step ensures that the invocation of every method within the app is captured in the logs;
\dcircle{3} The third phase is the most granular, involving the iteration of all statements in all methods of all classes. 
For each statement, a log statement is added immediately after the statement to be logged. 
This approach guarantees that if a statement is executed, it will be reflected in the log report. An example log entry would be: \texttt{LogCheckerClass.log("STATEMENT=com.\-example.My\-Class.foo()$\vert$\$r3 = (android.telephony.Telephony\-Mana\-ger) \$r2$\vert$4", "ANDROLOG")}. 
The number 4 is the line number that uniquely identifies the statement in the methods (this is needed since there might be similar statements in a given method); and
\dcircle{4} The final phase focuses on Android components such as Activities, Services, BroadcastReceivers, and ContentProviders. 
\al checks each class to determine if it is an Android component and, if so, adds a log statement in its lifecycle methods. 
For instance, in an Activity, the log would be added, not only, to the onCreate method, with a log statement like: \texttt{LogCheckerClass.log("ACTIVITY\-=\-c\-o\-m\-.\-example.My\-Activity", "ANDROLOG")} or \texttt{LogChecker\-Cl\-ass.\-log("SER\-VICE=com.\-example.MyService", "ANDROLOG")}.
After these phases are completed, the app is repackaged, aligned, and signed, making it ready for testing, manually or automatically.

\subsection{Log Processor Component}

\al's approach to computing code coverage after an app has been tested is divided into three main steps.
The process begins with \al generating a summary of the app. 
This summary includes quantifying the number of classes, methods, statements, and Android components in the app. 
This initial summary serves as a baseline for understanding the full scope of the app and computes the code coverage report.
The next step involves \al processing the log report previously generated during the app's execution. 
This log report, provided by the analyst, contains detailed information about the app elements that were executed. 
By parsing this report, \al identifies which classes, methods, statements, and components were actively used during the app's runtime.
In the final step, \al computes the code coverage by comparing the initial app summary with the execution summary derived from the log report. 
This comparison allows \al to determine the extent to which various app elements were executed.
By juxtaposing the execution scope (as outlined in the initial summary) with the actual execution data (from the log report), \al can quantify the coverage.
The outcome of this comparison is a code coverage report presented to the analyst.

\section{Future Research Opportunities}
\label{sec:future_research_question}

The introduction of \al creates opportunities for a variety of analytical and exploratory studies.
This section outlines two research opportunities that this tool can help address.

\noindent
\textbf{Comparative Effectiveness of Dynamic Analyses:}
\textit{How do existing dynamic analysis techniques, such as fuzzing, compare in terms of code coverage for Android applications?} 
This question uses \al to evaluate and contrast different dynamic analysis methodologies.
By assessing the code coverage achieved through these techniques, researchers can better understand their effectiveness in identifying, e.g., vulnerabilities or bugs in Android apps. This comparative analysis would be invaluable in refining dynamic analysis strategies.

\noindent
\textbf{Impact of Code Coverage on Defects Detection:}
\textit{What is the impact of code coverage on the detection of runtime behaviors and vulnerabilities?}
This question explores the relationship between the extent of code coverage in dynamic analysis and the effectiveness of identifying, e.g., bugs in apps.
Using \al to measure this coverage, researchers can gain insights into how thorough code exploration influences the detection of potential security and performance issues.
\section{Preliminary Empirical Evaluation}
\label{sec:evaluation}

In this section, we present preliminary empirical results that assess \al into instrumenting apps.
To that end, we have randomly selected \num{2000} recent apps (i.e., apps from 2023) from the AndroZoo dataset~\cite{10.1145/2901739.2903508}.
The average size of our dataset is 64 MB, and the median is 56 MB.
We performed our experiment on a Linux server (Debian 5.10.0-7-amd64) with an AMD EPYC 7552 48-Core Processor CPU with 96 cores and 630GB of RAM. 
Subsequently, for each app, we use \al to perform the instrumentation (with all levels of granularity activated), during which we also record the time taken for this instrumentation to complete.
\al was able to instrument 1957 apps successfully (i.e., 98\%).
The average instrumentation time is 34 seconds, and the median is 1 seconds.
The failure to successfully instrument 43 apps is attributed to the oversized DEX files generated by Soot, which require splitting for 17 apps. This issue is not a problem in general, except for pre-Lollipop Android apps (API 22).
For the remaining 26~apps, the error is due to a bug in Soot related to backward analysis on empty sets.

To compare \al against existing tools, we have considered COSMO, ACVTool, and BBoxTester, and performed the instrumentation step.
COSMO could instrument 1588 (79\%) apps, and ACVTool could instrument 966 (48\%) apps.
We encountered difficulties with BBoxTester, while the installation was successful, we faced challenges in the instrumentation process, as it was impossible to instrument any app due to many different crashes and errors. 
We attribute these problems to obsolescence.
All artifacts are available in the project's repository.
\section{Related Work}
\label{sec:related_work}

This section discusses existing research prototypes to instrument Android apps in a black-box manner.

InsDal~\cite{7884662}, CovDroid~\cite{7273401}, and the approach by Huang et al.~\cite{7226692} have been introduced as tools for computing code coverage in Android apps, but their unavailability to the public as open-source limits their practical application. 
InsDal, built on top of ApkTool, operates at the smali level, offering only instrumentation at the class and method levels. 
Similarly, CovDroid executes its instrumentation at the smali level, inserting probes at the method level.
The method devised by Huang et al. also functions at the smali level but necessitates additional modifications to the Android manifest, including integrating new permissions. 
Their tool demonstrates a limited success rate in instrumenting apps, achieving only 36\%.

BBoxTester, introduced in~\cite{7299958}, is a black-box code coverage tool for Android apps.
It converts Dalvik bytecode into Java bytecode using dex2jar, and then leverages Emma~\cite{emma} for the instrumentation process. 
However, BBoxTester's approach necessitates modifying the Android manifest and adding app resources, making it less flexible and potentially intrusive.
Similarly, COSMO~\cite{9438584} is designed to transform Dalvik bytecode into Java source code, subsequently using JaCoCo~\cite{jacoco} for instrumentation. 
An aspect of COSMO is its method of recording code coverage summaries directly on the device's SDCard.
ACVTool~\cite{10.1145/3395042} is another tool to instrument Android apps. 
Operating at the smali level, ACVTool's process requires several steps. 
This includes modifying the Android manifest to incorporate additional components and new permissions.
Furthermore, like COSMO, ACVTool records reports directly onto the device's SDCard. 
These procedures introduce potential intrusiveness.

\al, in contrast to the tools discussed, is an openly available Android app instrumentation framework that offers non-intru\-sive, granular-level analysis without requiring additional permissions or Android manifest modifications. 
\al achieves high instrumentation success with simple two-command-line usage. 
Built on the well-maintained Soot framework, \al ensures compatibility with newer apps and adaptability for future needs. 
This approach not only simplifies the instrumentation process but also minimizes interference with the original app functionality.
\section{Conclusion}
\label{sec:conclusion}

In this paper, we presented a novel framework, called \al, for automatically inserting logging probes into Android apps to compute code coverage in a black-box manner.
\al, built on top of the Soot framework, is designed to be \emph{simple} and to easily evolve on demand.
\al can insert probes at different levels of granularity such as classes, methods, statements, and Android components with a 98\% success rate.
Most importantly, \al adopts a streamlined approach to ensure non-interference with existing app code; it only adds log statements for behavior monitoring, avoiding new permissions, resources, Android components, or manifest modifications, thereby minimizing the risk of affecting the app's original functionality.
\section{Data Availability}
\label{sec:data_availability}

To promote transparency and facilitate reproducibility, we make \al and our artifacts available to the community at:
\begin{center}
\url{https://github.com/JordanSamhi/AndroLog}
\end{center}
We also provide a video demonstration of \al:
\begin{center}
\url{https://www.youtube.com/watch?v=NEcYg98k_o4}
\end{center}

\bibliographystyle{ACM-Reference-Format}
\bibliography{bib}

\end{document}